\documentclass[twocolumn,prl,aps,superbib,tightenlines,floatfix]{revtex4}
\usepackage{amsfonts,amsmath,amssymb,amsthm,bm}
\usepackage{graphicx}
\usepackage[dvipsnames,usenames]{color}
\begin{document}
\bibliographystyle{apsrev}
\title{Efficiency of Energy Conversion in Thermoelectric Nanojunctions}
\author{Yu-Shen Liu }
\author{Yi-Ren Chen}
\author{Yu-Chang Chen}
\email{yuchangchen@mail.nctu.edu.tw} \affiliation{Department of
Electrophysics, National Chiao Tung University, 1001 Ta Hsueh Road,
Hsinchu 30010, Taiwan }
\begin{abstract}

Using first-principles approaches, this study investigated the
efficiency of energy conversion in nanojunctions, described by the
thermoelectric figure of merit $ZT$. We obtained the qualitative and
quantitative descriptions for the dependence of $ZT$ on temperatures
and lengths. A characteristic
temperature: $T_{0}= \sqrt{\beta/\gamma(l)}$ was observed. When
$T\ll T_{0}$, $ZT\propto T^{2}$. When $T\gg T_{0}$, $ZT$ tends to a
saturation value. The dependence of $ZT$ on the wire length for
the metallic atomic chains
is opposite to that for the insulating molecules:
for aluminum atomic (conducting) wires, the saturation value
of $ZT$ increases as the length increases; while for alkanethiol
(insulating) chains, the saturation value of $ZT$ decreases as
the length increases. $ZT$ can also be enhanced by choosing
low-elasticity bridging materials or
creating poor thermal contacts in nanojunctions.

\end{abstract}
 \maketitle

There has been renewed interest in the study of
thermoelectricity motivated by its possible application in
energy-conversion devices at the nanoscale level
\cite{Paulsson,Zheng,Wang,Reddy,Reddy1,Galperin,
Pauly,Finch,Troels,Dubi,Liu}. Recent experiments on the Seebeck
coefficient, which is insensitive to the number
of molecules in the junction, shed light on the possibility of the
implementation of thermoelectric devices at the atomic level
\cite{Reddy}. Nanoscale energy-conversion devices
can convert waste heat energy into useful electric power and
stabilize miniature electronic devices by reducing the temperature.
The Seebeck coefficient, which is related not only to the
magnitude but also to the slope of density of states, can provide
more information than current-voltage characteristics \cite{Paulsson,Zheng,Wang,Reddy}.
Measurement of the Seebeck coefficient has been applied to explore the
effect of chemical structure on the electronic structure of
molecular junctions \cite{Reddy1}. The gate field has been
theoretically proposed as a means of modulating the conduction
mechanism between p-type [the Fermi energy is closer to the highest
occupied molecular orbital (HOMO)] and n-type [the Fermi energy is
closer to the lowest unoccupied molecular orbital (LUMO)] via the
sign of the Seebeck coefficient \cite{Wang,Liu}. Although much research
has been devoted to the study
of Seebeck coefficient, little is known about the efficiency of
energy conversion in nanojunctions \cite{Finch}. The objective of this
research was to provide greater insight into this subject.

Molecular tunneling junctions consist of source-drain electrodes as
independent electron and heat reservoirs with distinct
temperatures $[T_{L(R)}]$ and chemical potentials $[\mu _{L(R)}]$. The
efficiency of energy conversion depends on several factors: the electrical
conductance $(\sigma )$, the Seebeck coefficient $(S)$, the electron
thermal conductance $(\kappa _{el})$ and the phonon thermal
conductance $(\kappa _{ph})$. The efficiency can be described
by the dimensionless thermoelectric figure of
merit \cite{Troels}:

\begin{equation}
ZT=\frac{S^{2}\sigma }{\kappa _{el}+\kappa _{ph}}T,  \label{zt}
\end{equation}%
where $T=(T_{L}+T_{R})/2$ is the average temperature in the
source-drain electrodes. The ideal thermoelectric molecular junction
would have a large $S$, a large $\sigma $ and a small combined thermal
conductance $(\kappa_{el}+\kappa _{ph})$. Thermoelectric
materials with a large $\sigma $ are usually accompanied by a
large $\kappa _{el}$, which makes the enhancement
of the thermoelectric figure of merit a challenging task.

In this Letter, we reported first-principles calculations of the
thermoelectric figure of merit in nanojunctions. It aimed to obtain
a qualitative and quantitative descriptions of $ZT$ for temperatures
and lengths of the nanojunctions. The self-consistent density
functional theory (DFT) was performed together with the
derivation of an analytical expression for $ZT$ to investigate its
dependence on the temperatures and lengths of nanojunctions. As
an example, this study investigated $ZT$ for the aluminum
atomic (conducting) wires and the alkanethiol (insulating) molecules
in a nanojunction in the linear response regime. It was found
$ZT\varpropto T^{2}$ at low temperatures, while $ZT$ tended to 
a saturation value at high temperatures. The dependence of $ZT$ on
the wire lengths for the metallic atomic chains was opposite to that for the
insulating molecules: longer conducting wires and shorter insulating
molecules had better efficiency of energy conversion. The results of
this study may be of interest to experimentalists attempting to
develop thermoelectric nanoscale devices.

First, we started by a brief introduction of the DFT calculations for a
molecule sandwiched between two bulk electrodes with external source-drain
bias. The effective single-particle wave functions of the whole system were
calculated in scattering approach by solving the Lippmann-Schwinger equation
with exchange and correlation energy included within the local density
approximation iteratively until the self-consistency was
obtained. The effective single-particle wave
function $\Psi _{E}^{L(R)}(\mathrm{\mathbf{r},\mathbf{K}}_{||})$ represents
the electron incidents from the left
(right) electrode with the energy $E$ and component of the momentum $\mathrm{\mathbf{%
K}}_{||}$ parallel to the electrode surface \cite{Lang,Di Ventra3,Yang}.
These left- and right-moving wave functions, weighting with
the Fermi-Dirac distribution function according to their energies, were applied to
calculate the electric current $I$ and the thermal current conveyed by the
transport electrons $J_{Q}^{el}$ , via the following expressions: $I=\frac{2e%
}{h}\int {dE}\left[ {{f_{E}^{R}(\mu _{R},T_{R})\tau
^{R}(E)-f_{E}^{L}(\mu _{L},T_{L})\tau ^{L}(E)}}\right] $ and
$J_{Q}^{el}=\frac{2}{h}\int dE\left[ \left( E-\mu _{R}\right)
f_{E}^{R}\tau ^{R}(E)-\left( E-\mu _{L}\right) f_{E}^{L}\tau
^{L}(E)\right] $, where $\tau ^{L(R)}(E)$ is the transmission
function of the electron with energy $E$ incident from the left
(right) electrode. The transmission function can be computed using
the wave functions obtained self-consistently in DFT calculations
according to $\tau ^{L(R)}(E)=%
\frac{\pi \hbar ^{2}}{mi}\int {d\mathrm{\mathbf{R}}\int {d\mathrm{\mathbf{K}}%
_{||}}}I_{EE}^{LL(RR)}(\mathrm{\mathbf{r}},\mathrm{\mathbf{K}}_{||}),$
where $I_{E{E}^{\prime }}^{LL(RR)}=\left[ \Psi _{E}^{L(R)}\right]
^{\ast }\nabla \Psi _{{E}^{\prime }}^{L(R)}-\nabla \left[ \Psi
_{E}^{L(R)}\right] ^{\ast }\Psi _{{E}^{\prime }}^{L(R)}$ and
$d\mathrm{\mathbf{R}}$ represents an element of the electrode
surface. It is assumed that the left and right electrodes served as
independent electron and phonon reservoirs with
the electron population described by the Fermi-Dirac distribution function, $%
f_{E}^{L(R)}=1/\left( \exp \left( \left( E-\mu _{L(R)}\right)
/k_{B}T_{L(R)}\right)
+1\right) $, where $k_{B}$ is the Boltzmann constant, and $\mu _{L(R)}$ and $%
T_{L(R)}$ are the chemical potential and the temperature in the left (right)
electrode, respectively. The external source-drain bias is defined by: $%
V_{B}=(\mu _{R}-\mu _{L})/e$.

Then, we briefly described the method used to calculate the electrical
conductance, the Seebeck coefficient and the thermal conductance
conveyed by electron transport. 
We considered the extra electric and thermal current
induced by an additional infinitesimal temperature ($\Delta T$) and
voltage ($\Delta V$) symmetrically distributed across the junction:
\begin{widetext}
\begin{equation}
\Delta I=I(\mu _{L},T_{L}+\frac{\Delta T}{2};\mu _{R},T_{R}-\frac{\Delta T}{2%
})+I(\mu _{L}+\frac{e\Delta V}{2},T_{L};\mu _{R}-\frac{e\Delta V}{2}%
,T_{R})-2I(\mu _{L},T_{L};\mu _{R},T_{R}),\text{ }  \label{deltaI}
\end{equation}%
and
\begin{equation}
\Delta J_{Q}^{el}=J_{Q}^{el}(\mu _{L},T_{L}+\frac{\Delta T}{2};\mu _{R},T_{R}-%
\frac{\Delta T}{2})+J_{Q}^{el}(\mu _{L}+\frac{e\Delta V}{2},T_{L};\mu _{R}-\frac{%
e\Delta V}{2},T_{R})-2J_{Q}^{el}(\mu _{L},T_{L};\mu _{R},T_{R}),  \label{deltaJ}
\end{equation}%
\end{widetext}
respectively. After expanding the Fermi-Dirac distribution function to the
first order in $\Delta T$ and $\Delta V$, we obtained the Seebeck coefficient
(defined by $S=\Delta V/\Delta T$) by letting $\Delta I=0$ and the electron
thermal conductance (defined by $k_{el}=\Delta J_{Q}^{el}/\Delta T$):
\begin{equation}
S=-\frac{1}{e}\frac{\frac{K_{1}^{L}}{T_{L}}+\frac{K_{1}^{R}}{T_{R}}}{%
K_{0}^{L}+K_{0}^{R}},  \label{S}
\end{equation}
\begin{equation}
\kappa _{el}=\frac{1}{h}\sum_{i=L,R}[K_{1}^{i}eS+\frac{K_{2}^{i}}{T_{i}}],
\label{kel}
\end{equation}
where $K_{n}^{L(R)}=-\int dE\left( E-\mu _{L(R)}\right) ^{n}\frac{\partial
f_{E}^{L(R)}}{\partial E}\tau (E)$, and $\tau (E)=\tau ^{R}(E)=\tau ^{L}(E)$%
, a direct consequence of the time-reversal symmetry. In addition, the
differential conductance, typically insensitive to temperature in cases
where direct tunneling is the major transport mechanism, may be expressed as:
\begin{equation}
\sigma =\frac{e}{2}\int \sum_{i=L,R}\frac{f_{E}^{i}(1-f_{E}^{i})}{k_{B}T_{i}}%
\tau (E)dE.  \label{sigma}
\end{equation}
So far, the physical quantities that have been discussed have been related
to the propagation of electrons. However, in most cases, the thermal
current is dominated by the contribution from phonon transport.
In the absence of the phonon thermal conductance, the research
on $ZT$ is incomplete. To consider the phonon contribution to $ZT$, it is
assumed that the nanojunction is a weak elastic link, with a given stiffness that
may be evaluated from total energy calculations, attached to the
electrodes modeled as phonon reservoirs. We estimate the contribution
of the thermal current from phonon scattering ($J_{Q}^{ph})$, following the approach
of Patthon and Geller \cite{Patton}. After expanding the
Bose-Einstein distribution function to the first order of $\Delta T$
in the expression of phonon thermal current, the phonon
thermal conductance(defined by $k_{ph}=\Delta J_{Q}^{ph}/\Delta T$) is obtained:
\begin{equation}
\kappa _{ph}=\frac{\pi K^{2}}{\hbar k_{B}}\int
dEE^{2}N_{L}(E)N_{R}(E)\sum_{i=L,R}\frac{n_{i}(E)(1+n_{i}(E))}{T_{i}^{2}},
\label{kph}
\end{equation}%
where $n_{L(R)}\equiv 1/(e^{E/K_{B}T_{L(R)}}-1)$ and
$N_{L(R)}(E)\simeq CE$ is the Bose-Einstein distribution function
and the spectral density of phonon states in the left (right)
electrode, respectively. The stiffness of the bridging nano-structure
is: $K=YA/l$, where $Y$ is the Young's modulus and $A$
$(l)$ is its cross-section (length).

Finally, $ZT$ could be calculated by applying
Eqs.~$(\ref{S})$ to $(\ref{kph})$. The
Seebeck coefficient and the electron (phonon) thermal
conductance can be characterized by the power law expansions: $S\approx
\alpha T$, $\kappa _{el}\simeq \beta \left[ T+\eta T^{3}\right] \approx
\beta T$ and $\kappa _{ph}=\gamma (l)T^{3}$ in the common range of
temperatures ($T_{L}\approx T_{R}=T$) and in the linear response regime ($%
\mu _{L}\approx \mu _{R}=\mu $), where $\alpha =-\pi ^{2}k_{B}^{2}\frac{%
\partial \tau (\mu )}{\partial E}/\left( 3e\tau (\mu )\right) $; $\beta
=2\pi ^{2}k_{B}^{2}\tau (\mu )/(3h)$; $\eta =(\pi k_{B}\partial \tau
(\mu )/\partial E)^{2}/\left( 3\tau (\mu )^{2}\right) $ and $\gamma
(l)=8\pi ^{5}k_{B}^{4}C^{2}A^{2}Y^{2}/(15\hbar l^{2})$.
Consequently, the thermoelectric figure of merit in the
nanojunctions has a simple form,

\begin{equation}
ZT\approx \frac{\alpha ^{2}\sigma T^{3}}{\beta T+\gamma (l)T^{3}},
\label{ZT2}
\end{equation}
which is valid in small bias and low temperature regimes.

\begin{figure}
\includegraphics[width=7.5cm,height=8.50cm]{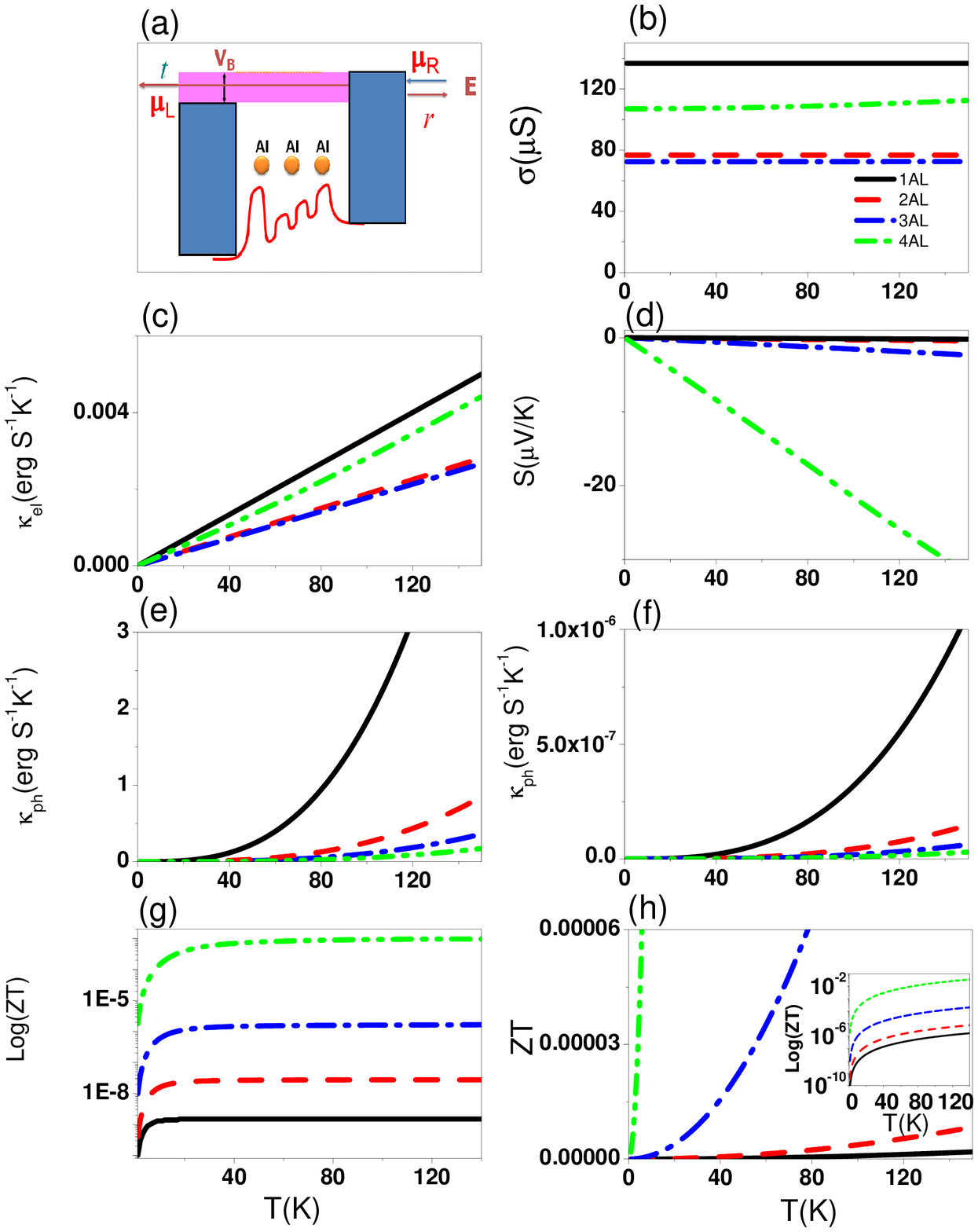}
\caption{Aluminum atomic junctions at $V_{B
}=0.01$~V: (a) Schematic of 3-Al atomic chain and its Energy
diagram. The Al-Al bond distance was about $6.3$~a.u.; (b) Electrical conductances $\sigma$ vs T; (c)  Electron
thermal conductances $\kappa_{el}$ vs T; (d) Seebeck coefficients
$S$ vs T; (e) Phonon thermal conductances $\kappa_{ph}$ vs T
($Y=1.2\times 10^{13}$~dyne/cm$^{2}$);  (f) Phonon thermal
conductances $\kappa_{ph}$ vs T ($Y=5.0\times 10^{9}$~dyne/cm$^{2}$);
(g) $Log(ZT)$ vs T (for $Y=1.2\times 10^{13}$~dyne/cm$^{2}$);
(h) $ZT$ and $Log(ZT)$ (Inset) vs T (for $Y=5.0\times
10^{9}$~dyne/cm$^{2}$).}
\label{Fig1}
\end{figure}
\begin{figure}
\includegraphics[width=8.0cm,height=8.500cm]{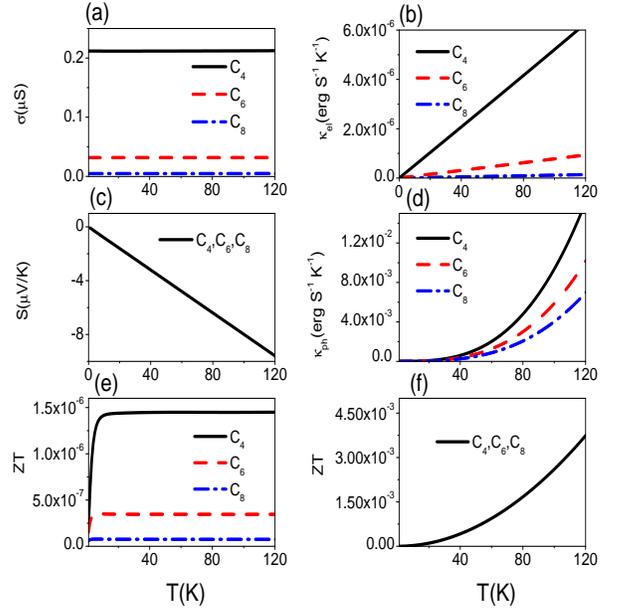}
\caption{Alkanethiol junctions at $V_{B}=0.01$~V:
(a) Electric conductance $\sigma$ vs T; (b) Electron thermal
conductances $\kappa_{el}$ vs T; (c) Seebeck coefficients $S$ vs T;
(d) Phonon thermal conductance $\kappa_{ph}$ vs T 
($Y\simeq 2.3\times 10^{12}$~dyne/cm$^{2}$); 
(e) $ZT$ vs T ($Y\simeq 2.3\times 10^{12}$~dyne/cm$^{2}$); 
(f) $ZT$ vs T (for $Y=0$~dyne/cm$^{2}$).}
\label{Fig2}
\end{figure}

The properties of the thermoelectric figure of merit now can be discussed
using Eq.~$\left( \ref{ZT2}\right) $. There was a
characteristic temperature, $T_{0}\equiv \sqrt{\beta /\gamma (l)}$,
for $ZT$ in the nanojunctions. When $T\ll T_{0}$, the thermal
current was dominated by the
contribution from the electron transport ($k_{ph}\ll k_{el}$), which led to $%
ZT$ increasing as the temperature increased: $ZT\approx \sigma
S^{2}T/k_{el}\approx \left[ \alpha ^{2}\sigma /\beta \right] T^{2}$.
Similarly, when $T\gg T_{0}$, the thermal current was dominated by the
contribution from the phonon transport ($k_{ph}\gg k_{el}$), which led to a
saturation of $ZT$ at a constant value related to the length of the
junction: $ZT\approx \sigma S^{2}T/k_{p}\approx \alpha ^{2}\sigma /\gamma
(l)$. 
To increase $ZT$ it was
first necessary to reduce $k_{ph}$ by choosing low-elasticity bridging
wires or creating poor thermal contacts in the nanojunctions, such that $%
ZT\approx \sigma S^{2}T/k_{el}$. It is worth noting that $\sigma $ and $k_{el}$
roughly canceled each other out in the contribution of $\ ZT$ because both were
proportional to $\tau (\mu )$. It then followed that $ZT\varpropto S^{2}T$ and,
thus, that the material with a large Seebeck coefficient was of key importance to
increasing $ZT$. The characteristic mark of such a material in the
nanojunctions is a sharp peak around the Fermi levels in the DOS \cite{Liu}, and
the Seebeck coefficient may be optimized by applying the gate field \cite%
{Wang,Liu}. In addition, it was noted that $\alpha $ and $%
\sigma $ depend on the length of the junction in a way related to the
material properties of bridging wires, which is reflected in the
distinguished features of $ZT$ on the length dependence. This
point was explained using two catalogs of nanojunctions: the aluminum
atomic (conducting) wires and the alkanethiol (insulating) chains, as
discussed below.

Aluminum atomic wire is ideal for studying charge
transport at the atom-scale [see Fig.~\ref{Fig1}(a) for a schematic
of the aluminum junction] \cite{Lang1,Koba,Yang1,Cuevas}. As shown
in Fig.~\ref{Fig1}(b), the conductance was relatively insensitive to the chain
length (typically around $1~G_{0}=2e^{2}/h\approx 77~\mu S$) apart
from the possible 4-atom periodicity due to a filling factor
of $1/4$ in the $\pi$ orbitals \cite{Thygesen}. As shown in
Fig.~\ref{Fig1}(c), the magnitude of electron
thermal conductance was linear in temperatures, $\kappa _{el}\approx
\beta T$. At a fixed temperature, the dependence of the magnitude of
$\kappa _{el}$ on the number of Al atoms was
the same as that of $\sigma $, owing to the fact that both $\sigma $ and $%
\kappa _{el}$ were proportional to $\tau (\mu )$. As shown in
Fig.~\ref{Fig1}(d), the magnitude of the Seebeck coefficient was linear in
temperature, $S\approx \alpha T$, with the negative sign showing that
the carrier was n-type. At a fixed temperature, it was observed that the
magnitude of the Seebeck coefficient increased considerably as the number
of Al atoms increased. The increase of the Seebeck coefficient was
due to the increase of the slope in the DOS at the Fermi level. These
features may be related to the fact that the Fermi level was close to
the LUMO in the Al wires. Fig.~\ref{Fig1}(e) shows the phonon thermal
conductance: $\kappa _{ph}=\gamma (l)T^{3}$, for the Young modulus
using $Y=1.2\times~10^{13}$~dyne/cm$^{2}$ from the total energy
calculations \cite{parametersAlatoms}. As seen, $\kappa _{ph}\gg \kappa _{el}$
was due to the large Young modulus.
Fig.~\ref{Fig1}(g) shows the thermoelectric figure of merit
with $\kappa _{ph}$ calculated
using $Y=1.2\times~10^{13}$~dyne/cm$^{2}$ from the total energy
calculations. The increase in the number of Al atoms sharply
increased the saturation value of $ZT$ because of the sharp increase
in the Seebeck coefficient
by the number of Al atoms according to $ZT\propto S^{2}$. The
thermoelectric figure of merit reached the saturation value,
$ZT\rightarrow \alpha ^{2}\sigma /\gamma (l)$ when $T\gg T_{0}$.
Since the mechanical elasticity of the Al wires
could be delicate to the detailed geometry in the contact region which
was unknown in the real experiment, as Fig.~\ref{Fig1}(f) shows the
phonon thermal conductance $\kappa _{ph}$ for another possible
value, $Y=5.0\times 10^{9}$~dyne/cm$^{2}$ from controlled tensile
experiments on nanoscale Al films \cite{Haque}. As shown
in Fig.~\ref{Fig1}(h), it is worth noting that $ZT$ could
be strongly enhanced by a smaller $\kappa _{ph}$.
In such cases, the thermal current conveyed by electron transport
dominated so that
$ZT\approx \left( \alpha ^{2}\sigma /\beta \right) T^{2}$
and $ZT$ was strongly enhanced.

Alkanethiols [CH$_{3}$(CH$_{2}$)$_{n-1}$SH, denoted as C$_{n}$] are a good
example of reproducible junctions that can be fabricated \cite{WangW,Tao1}.
In contrast to the conductor behavior of aluminum wires, alkanethiol
chains are insulators. It has been established that non-resonant tunneling
is the main conduction mechanism in alkanethiol junctions. Consequently,
the conductance is small and decreases exponentially with the length
of wire, as $\sigma =\sigma _{0}\exp \left( -\xi l\right) $ where $l$ is
the length of alkanethiol chain and $\xi \approx 0.78$ \r{A}$^{-1}$
\cite{Wold1,Beebe,Zhao,Kaun,Ma}, as shown in Fig.~\ref{Fig2}(a). By
exploiting the periodicity in the $\left( CH_{2}\right) _{2}$ group of the
alkanethiol chains, the wave functions of the C$_{n}$ junctions
were calculated by a simple scaling argument, which led to
exponential scaling in the transmission function $\tau (E)$. As
shown in Fig.~\ref{Fig2}(b), the magnitude of electron thermal
conductance was linear in temperatures, $\kappa _{el}\approx \beta T$.
At a fixed temperature, the magnitude of $\kappa _{el}$ decreased
exponentially with $n$, the number of carbon atoms in C$n$, owing
to the scaling behavior of $\tau (E)$. As shown in Fig.~\ref{Fig2}(c),
the magnitude of the Seebeck coefficient was linear in temperature
as $S\approx -\pi ^{2}k_{B}^{2}\frac{%
\partial \tau (\mu )}{\partial E}/\left( 3e\tau (\mu )\right) T$, and its
dependence on the number of carbon atoms was canceled due to the same scaling
factor $\exp \left( -\xi l\right) $ for both $\tau (\mu )$ and $\frac{%
\partial \tau (\mu )}{\partial E}$. As shown in Fig.~\ref{Fig2}(c),
thermal conductance increased as the temperature increased
as $\kappa_{ph}=\gamma (l)T^{3}$ for the Young modulus calculated with
total energy calculations \cite{parametersAlatoms}. At a fixed
temperature, $\kappa _{ph}$ decreased as $n^{-2}$ due to $\gamma
(l)\varpropto l^{-2}$ (see Fig.~\ref{Fig2}(d)). Due to the small transmission
probability for the insulating alkanethiol chains, the electron thermal
conductance (note: $k_{el}\varpropto \sigma $) was much suppressed
so that $\kappa _{el}\ll k_{ph}$, as shown in Fig.~\ref{Fig2}(b) and (d).
Consequently, the characteristic temperature $T_{0}$ was low in the
alkanethiol chains, and the $T^{2}$ regime for $ZT$ was significantly
suppressed. As shown in Fig.~\ref{Fig2}(e). $ZT$ decreased as the number
of carbon atoms increased for $T\gg T_{0}$,
due to the saturation value of $ZT\approx \alpha ^{2}\sigma /\gamma
(l)\varpropto l^{2}\exp \left( -\xi l\right) $. Nevertheless, there
was enough experimental evidence to show that the junctions had
poor thermal contacts for certain samples \cite{Smit}. These
samples quickly frustrated at much smaller biases. The frustration
of these samples could have been due to poor heat dissipation by the thermal
current via phonon transport when the local heating was triggered by
an external bias larger than the threshold value
\cite{Yang1,Chen1}. In such cases, the bridging nano-structure
effectively has a very small Young modulus. In the limit of
extremely poor thermal contacts (effectively,
$k_{ph}=0$, $\sigma $ and $k_{el}$ canceled the length dependence),
which led to $ZT\varpropto \alpha^{2}\sigma_{0}T^{2}/\beta $ and
implied that $ZT$ was independent of the wire length as shown in
Fig.~\ref{Fig2}(g).

In conclusion, self-consistent DFT calculations were performed to
study the efficiency of energy conversion in nanojunctions. There was a
characteristic temperature of $T_{0}$ for $ZT$: when $T\ll T_{0}$, $%
ZT\rightarrow \left( \alpha ^{2}\sigma /\beta \right) T^{2}$; when
$T\gg T_{0}$, $ZT\rightarrow \alpha ^{2}\sigma /\gamma (l)$. Of
key importance to increasing the efficiency of energy conversion
was using materials with a large Seebeck coefficient. Such materials were
usually characterized by a sharp peak around the Fermi levels in the
DOS. Efficiency could be further optimized by applying the gate
field, choosing low-elasticity bridging materials or creating poor
thermal contacts in nanojunctions. The relation between $ZT$ and the wire lengths
depended on the material properties: for aluminum atomic
(conducting) wires, the saturation value of $ZT$ increased as the
length increased; while for the alkanethiol (insulating) chains, the saturation
value of $ZT$ decreased as the length increased. The conclusions of
this study may be beneficial to research attempting to increase the
efficiency of energy conversion in nano thermoelectric devices.

The authors thank MOE ATU, NCTS and NCHC for support under Grants
NSC 97-2112-M-009-011-MY3 , 097-2816-M-009-004 and 97-2120-M-009-005.


\begin{thebibliography}{99}
\bibitem{parametersAlatoms}  The aluminum (alkanethiol) junction consisted of an Al atomic chain (a C$_{4}$ molecule) sandwich between two Al (Au) electrodes that were modeled as electron jellium with $r_{s}\approx 2$ ($r_{s}\approx 3$). The
following parameters were used for the junctions: the effective
cross-section chosen was $A=23.1$~a.u.$^{2}$ ($21.4$~a.u.$^{2}$) for the aluminum (alkanetiol) junction. The spectral densities of electrodes were evaluated using the longitudinal and transverse
sound velocities for Al (Au), $v_{l}=6.35~(3.2)\times 10^{5}$~cm/s and $v_{t}=3.1~(1.2)\times
10^{5}$~cm/s, respectively. The Young's
modulus of aluminim (alkanethiol) wire, $Y\simeq 1.2\times 10^{13}$~dyne/cm$^{2}$ ($Y\simeq 2.3\times 10^{12}$~dyne/cm$^{2}$), was calculated with total
energy calculations and was found to be almost independent of the length of the wires.
\end{thebibliography}
\end{document}